\documentclass[12pt]{article}
\usepackage{epsfig}
\newcounter{nfig}

\newcommand \ds \displaystyle

\begin{document}

\title{Electromagnetic Pion and Nucleon Form Factors at Positive and Negative
$q^2$ Within the Framework of Quark-Gluon Strings Model}
\date{The Date }

\author{A.B. Kaidalov $^1$, L.A. Kondratyuk$^1$ and D.V. Tchekin $^2$}

\maketitle

$^1${\it \ Institute for Theoretical and Experimental Physics, 117259, Russia} 

$^2$ {\it Moscow Institute of Physics and Technology, Dolgoprudny, Russia}

\begin{abstract}\label{abstr}
The electromagnetic form factors of pion and nucleon are considered within the 
framework  of the Quark-Gluon Strings Model, where the dependence of the form 
factors  on $q^2$ is determined by the intercept of a dominant Regge trajectory 
and Sudakov  form factor. The analytical expressions for the form factors found 
in the time-like region can be directly continued to the space-like region.
Good agreement with available experimental data on the pion and magnetic
nucleon  form  factor is obtained at positive as well as negative $q^2$. It is 
shown that the difference in $F_\pi$ and $G_m$ at positive and negative values 
of $q^2$ is  mainly  related to the behavior of the double logarithmic term in
the  exponent of  Sudakov form factor.  The model describes also existing
data  on the Pauli nucleon form factor $F_2$ and the ratio $G_e/G_m$.
\end{abstract}

\section{ INTRODUCTION}\label{intro}

It is well known that at large $\left|q^2\right| \gg M^2$ the charge and
magnetic nucleon form factors can be approximated quite well by the dipole
formula: $\left| G_{e,m}\right| \sim 1/\left|q^2\right|^2$ (see review
\cite{Stoler}). For Pauli form factors one has  $\left| F_1\right| \sim 1/
\left|q^2 \right|^2$, $\left| F_2\right| \sim 1/\left| q^2\right|^3 $.
Experimental data \cite{Baldini,Gauzzi} suggest also that at large $\left|
q^2\right| $ the proton magnetic form factor $G_m(q^2)$ in the time-like
region is approximately twice as large as that in the space-like region: 
$\left| G_m\left( q^2\right) \right| \sim c_1/\left| q^2\right|^2 $ for
$q^2>0$,  $\left| G_m\left( q^2\right) \right| \sim c_2/\left| q^2\right|^2 $
for $q^2<0$ , $c_1\simeq 2c_2$ (see e.g. review \cite{Gauzzi} and references
therein).

Available experimental data on the pion form factor can also be described
by the power law: $F_\pi \sim c_\pi/q^2$. There are also indications that the
modulus of the pionic form factor $\vert F_\pi \left( q^2 \right) \vert$ in
the time-like region is approximately twice as large as that in the space-like
region \cite{Gauzzi}.

As concerning theoretical development, there is a consensus that the correct 
description of form factors at asymptotically large $q^2$ is given by the
Hard-Scattering Picture (HSP) (see \cite{Brodsky} and references therein).
However, there is a question whether or not HSP based on perturbative QCD
can be applied to the description of the available data
\cite{Ioffe:1983qb,Isgur,Radyushkin,Aznaurian:1992tb}. Last years important
progress in modifying HSP via the inclusion of the intrinsic ${\bf
k}_{\perp}$-dependence of the wave function and sophisticated parametrizations
ofSudakov form factor has been made (see \cite {Botts,Li1,Li2,Kroll1}). This
new approach allowed to calculate the perturbative contribution to form
factors in a self-consistent way even at moderate momentum transfers (about
2-3 GeV) and to demonstrate that the perturbative contributions are too small
as compared with available data (see \cite{Boltz,Hyer,Kroll2}). Analysis of
the most recent developments in perturbative QCD calculations of hadronic
electromagnetic form factors (see review \cite{Jain}) demonstrated that at
least the normalization of the form factor cannot be predicted reliably by a
leading order calculations in $\alpha_s$. This result means that substantial 
nonperturbative (soft) contributions to the form factors can really be present.

A model of soft contributions to form factors based on the Quark-Gluon
Strings Model (QGSM) was proposed in \cite{Kaidalov1}. Previously QGSM was
used in refs.\cite{Kaidalov2,Kaidalov3} to describe soft hadronic
interactions at high energies. It is based on the ideas of  $1/N$ expansion
\cite{t'Hooft,Veneziano,Chang,Chew} and color tube model 
\cite{Casher1,Artru,Casher2,Gurvich}. QGSM can be considered as a microscopic
model for Regge phenomenology, and it makes it possible to relate many
different soft hadronic reactions. Within the framework of QGSM the
$q^2$-dependence of the charge and magnetic nucleon form factors as well as
the pionic form factor can be described in terms of the intercept of a Regge
trajectory and Sudakov form factor  \cite{Kaidalov1}.

In this paper we show that the differencies in moduli of hadronic form factors 
in the time-like and space-like regions can naturally be explained by analytical
dependence of Sudakov form factor on $q^2$. Originally QGSM was
formulated in terms of probabilities for quark-hadron and hadron-quark
transitions in the impact parameter representation. Here we generalize the
model introducing spin-dependent amplitudes for those transitions and
developing  a method for treatment of spin variables in QGSM. Earlier spin
effects in QGSM were discussed in \cite{Grigorian} and \cite{Kaidalov1}. Our
treatment of spin effects is essentially different from the approach used in
\cite{Grigorian}. Using our approach we separated the nucleon form factors
$F_1$ and $F_2$ and demonstrated additional suppression of the pion form
factor at large $q^2$, which is caused by chirality conservation.

As the answers for all the form factors are written analytically, they can
be continued  from the time-like region to the space-like region. Different
values of the form factors at positive and negative $q^2$ are accounted for by
the analytical behavior of the doubly logarithmic term in the exponent of
Sudakov form factor.

The paper is organized as follows. In Section \ref{model} the basic concepts
of QGSM are introduced. We define the amplitudes for quark-antiquark
transitions to nucleon-antinucleon and pion-antipion pairs: 
$A^{q\overline{q}\rightarrow N \overline{N}}\left( s,t\right) $, 
$A^{q\overline{q}\rightarrow \pi \overline{\pi}}\left( s,t\right) $  and
derive their asymptotic expressions at large $s$ and finite $t$. Then we write
 analytical formulas for $\gamma \rightarrow N\overline{N}$ and  $\gamma
\rightarrow \pi \overline{\pi}$ matrix elements at $\left| s\right| \gg M^2$,
$\left| t\right| \leq M^2$ expressing them as a convolution of the transition
amplitudes $\gamma \to q {\bar q}$ and $q {\bar q} \to h {\bar h}$ in momentum
representation or as  a product in impact parameter representation. In Section
\ref{spin}, we discuss the spin structures of $ A^{q\overline{q}\rightarrow
N\overline{N}}$, $A^{q\overline{q}\rightarrow \pi \overline{\pi}}$ amplitudes,
and derive expressions for the pion form factor $F_{\pi}(s)$ and for the
magnetic and charge nucleon form factors $G_m(s)$, $G_e(s)$. Then, in Section
\ref{sudak} we consider additional suppression of the matrix elements  
$\gamma \rightarrow N\overline{N}$ and $\gamma \rightarrow \pi \overline{\pi}$
caused by the Sudakov form factor. In Section \ref{exper} we present the
results of numerical calculations for $G_m(s)$, $G_e(s)$ and  $F_{\pi}$ in the
space-like and  time-like regions and compare them with the available
experimental data on the pion and nucleon form factors. Our conclusions are
presented in Section \ref{concl}.

\section{ Quark-Gluon Strings Model. Transition amplitudes $T^{q\overline{q}
\rightarrow h\overline{h} }\left( s,t\right) $ in the limit of large $s$ and
finite $t$.}\label{model}

Let us consider binary reactions  $\pi ^{+}\pi ^{-}\to \pi^0  \pi ^0$, $\pi
^{+} \pi^{-}\to N \overline{N}$ and  $p \overline{p}\to N\overline{N}$ which
at large $s$ and finite $t$ can be described by the planar graphs with valence
quark exchanges in $t-$ channel (see diagrams  a)-c) in Fig.1). Note that
solid lines in diagrams of Fig.1 correspond to valence quarks, and soft gluon
exchanges are not shown. According to the topological $1/N_f$ expansion (TE) 
\cite{Kaidalov1,Veneziano} those graphs give dominant contributions to the
corresponding amplitudes when $N_f\gg 1$ and $N_c/N_f\sim 1$. In the cases
considered here the exchanges of light quarks $u,d,s$ are mainly important and
the parameter of expansion seems not to be very small $1/N_f=1/3$. However,
for the amplitudes with definite quantum numbers in the $t$-channel the actual
parameter of expansion is $1/N_f^2$ \cite{Veneziano}. Each graph of TE has a
rather simple interpretation within the framework of space-time picture which
can be formulated using color tube (or color string) model \cite{Kaidalov4}.

As an example we consider a space-time picture of the reaction $\pi ^{+}\pi
^{-}\to \pi^0\overline{\pi }^0$ which correspond to the graph of Fig.1a). At
high energy $\sqrt{s}$ this reaction occurs due to a  rare quark-parton
configuration in each pion when in the c.m.s. a spectator  quark (or
antiquark) takes almost all the momentum of hadron and the other (valence)
antiquark (or quark) is rather slow. The difference in rapidity between the
quark and antiquark in each pion is

\begin{equation}
\label{rapidity}y_q-y_{\bar q}\simeq \frac12\ln \left( \frac s{s_0}\right) 
\end{equation}
where $s_0 \simeq 1$ GeV$^2$.
Then two slow quarks from $\pi ^{+}$ and $\pi ^{-}$ annihilate, and the fast
spectator quark and antiquark continue to move in the previous directions and
produce a color string in the intermediate state. Then the string breaks due
to the production of a $q\bar q$-pair from the vacuum creating a $\pi ^0
\overline{\pi}^0$ -system in the final state. The same space-time picture
holds for the graph of Fig.1b) with only difference that the string breaks 
producing a diquark-antidiquark pair from the vacuum and creating a $N\bar N
$-system  in the final state.

Correspondingly the graph of Fig.1c) shows the formation of the
$q\overline{q}$ string due to annihilation of the valence  diquark-antidiquark
pair in the initial state and production of another diquark-antidiquark pair
due to the break of the string. Annihilation of the initial $q\overline{q}$
(or $qq\overline{qq}$) pair takes place when the difference in  rapidity of the
valence $q$ and $\overline{q}$ (or $\overline{qq}$) is small (both interacting
partons are almost at rest in c.m.s.) and relative impact parameter $\vert
{\bf  b}_{\perp} -{\bf b}_{0\perp} \vert$ is less than the interaction radius.
This can be described by the probability to find a valence quark with
rapidity $y_q$ and impact parameter ${\bf b}_{\perp }$ inside a hadron has
the following form \cite{Kaidalov1}

\begin{equation}
\label{probability}w\left(y_q-y_0,{\bf b}_{\perp}-{\bf b}_{0\perp}\right)=
\frac c{4\pi R^2(s)}\exp \left[ 
-\beta (y_q-y_0)-\frac{({\bf b}_{\perp}-{\bf b}_{0\perp})^2}{4R^2(s)}\right]
\end{equation}

As it was shown in \cite{Kaidalov1} it is possible to relate the parameters
$\beta $ and $R^2(s)$ of the quark distribution inside a hadron to the
phenomenological parameters of a Regge trajectory $\alpha _i (t)$ which gives 
dominant contribution to the amplitude corresponding to the considered planar
graph. In this case $R^2(s)=R_0^2+\alpha ^{\prime }(y_q-y_0)$ is the
effective interaction radius squared, $y_0$ is the average rapidity, ${\bf
b}_{\perp}$ is the transverse coordinate of the c.m. system in the impact
parameter representation, $\alpha ^{\prime }=\alpha _R^{\prime }(0)$ is the
slope of the dominant Regge trajectory, and $\beta $ is related to the
intercept as $\beta =1-\alpha_R(0)$.

Due to creation of a string in the intermediate state the amplitude of a
binary reaction $ab\to cd$ has the $s$-channel factorization property: the
probability for the string to produce different hadrons in the final state
does not depend on the type of the annihilated quarks and is only determined
by the types of the produced quarks. The same can be said about the process
of the production of the color string in the intermediate state from the
initial hadron configuration: this process depends only on the type of the
annihilated quarks. This $s$-channel factorization was formulated in \cite
{Kaidalov2},\cite{Kaidalov3} in terms of probabilities defined by
eq.(\ref{probability}).

Generalizing this approach let us introduce amplitudes
$\widetilde{T}^{ab\to q\bar q} (s,{\bf  b}_{\perp})$ and
$\widetilde{T}^{q\bar q\to cd}(s,{\bf b}_{\perp })$ that describe a
formation and a fission of the intermediate string respectively. The amplitude
of the binary reaction $ab \to cd$ described by the planar graph of Fig.1a) (
b) or c)) can be written employing the $s$-channel factorization property as
the convolution of two amplitudes

\begin{equation}\label{ImpRepr}
A^{ab\rightarrow cd}\left(
s,{\bf q}_{\perp }\right) = \frac i{8\pi ^2s}\int d^2{\bf k}_{\perp } \ 
T^{ab\rightarrow q\overline{q}} \left( s,{\bf k }_{\perp }\right) 
T^{q\overline{q}\rightarrow cd}\left( s,{\bf  q }_{\perp }-{\bf
k}_{\perp }\right)
\end{equation}
in the momentum representation, or as the product

\begin{equation}\label{factorization}
\widetilde{A}^{ab\to cd}(s,{\bf b}_{\perp })=\frac
i{2s}\ \widetilde{T}^{ab\to q\bar q}(s,{\bf b}_{\perp })\ \widetilde{T}
^{q\bar q\to cd}(s,{\bf b}_{\perp }) 
\end{equation}
in the impact parameter representation.

Let us find the solution for the quark-hadron transition amplitudes $T^{\pi 
^{+}\pi ^{-}\rightarrow q
\overline{q}}\left( s,{\bf k}_{\perp }\right)$ and $T^{q   
\overline{q}\rightarrow N\overline{N}}\left( s,{\bf
k}_{\perp }\right) $ at large $s$, which corresponds to the single Regge-pole
parameterizations  of the binary hadronic amplitudes $A^{\pi ^{+}\pi ^{-
}\rightarrow \pi^0
\pi^0}$,  $A^{\pi \overline{\pi }\rightarrow N\overline{N}}$ and
$A^{N\overline{N}\rightarrow N\overline{N}}$:

\begin{equation} \label{MBDReggeAmplitude}
\begin{array}{c}
\displaystyle A^{\pi ^{+}\pi ^{-}\rightarrow \pi
^0\pi ^0}\left( s,t\right) =N_M\ \left( -\frac
s{m_0^2}\right) ^{\alpha _M\left( t\right) }\exp \left( R_{0M}^2t\right) \\ \\
\ds A^{\pi \overline{\pi }\rightarrow N 
\overline{N}}\left( s,t\right) =N_B\ \left( -\frac
s{m_0^2}\right) ^{\alpha _B\left( t\right) }\exp \left( R_{0B}^2t\right) \\ \\
\ds A^{N\overline{N}\rightarrow
N\overline{ N }}\left( s,t\right) =N_D\ \left( -\frac
s{m_0^2}\right) ^{\alpha _D\left( t\right) }\exp \left( R_{0D}^2t\right) 
\end{array}
\end{equation}

Here $\alpha _M\left( t\right) $, $\alpha _B\left( t\right) $ and
$\alpha_D\left( t\right) $ are the dominant meson, baryon and
diquark-antidiquark trajectories; $N_M$, $N_M$ and $N_D$ are the normalization
constants. We have the following intercepts and slopes for the dominant Regge
trajectories

\begin{equation}
\label{ReggePoles}\alpha _M\left( 0\right) \simeq 0.5,\quad \alpha _B\left(
0\right) \simeq -0.5,\quad \alpha _D\left( 0\right) \simeq -1.5
\end{equation}
and

\begin{equation}\label{ReggePoles2}
\alpha _M^{\prime }\left( 0\right) \simeq \alpha _B^{\prime }\left( 0\right)
\simeq \alpha _D^{\prime }\left( 0\right) \simeq 1.0\ GeV^{-2} .
\end{equation}

Taking into account equations (\ref{rapidity}), (\ref{MBDReggeAmplitude}) we
can write  the amplitudes  $\widetilde{T}^{q\overline{q} \rightarrow \pi
\overline{\pi}}\left( s,{\bf b}_{\perp }\right) $ and
$\widetilde{T}^{q \overline{q}\rightarrow N\overline{N}}\left( s,{\bf b
}_{\perp }\right)$ as follows

\begin{equation} \label{PionNucleonFragmentation}
\begin{array}{c}
\ds \widetilde{T}^{q\overline{q} \rightarrow \pi \overline{\pi
}}(s,  {\bf b}_{\perp})=N_M^{1/2}\frac 1{2\sqrt{\pi }R_M\left( s\right)
}\left( -\frac s{m_0^2}\right) ^{\left( \alpha _M\left( 0\right) +1\right)
/2}\exp \left( -  \frac{{\bf b}_{\perp }^2}{8R_M^2\left( s\right) }\right) \\
\\ 
\ds \widetilde{T}^{q\overline{q}\rightarrow N \overline{N}}(s,{\bf
b}_{\perp})= N_D^{1/2}\frac 1{2\sqrt{\pi }R_D\left(s\right) }
\left( -\frac s{m_0^2}\right)^{\left( \alpha _D\left( 0\right)+1\right) /2}
\exp \left( -\frac{{\bf b}_{\perp }^2}{8R_D^2\left( s\right)}\right)
\end{array}
\end{equation}
where $R_M\left( s\right) $ and $R_D\left( s\right) $ are the effective
interaction radii

\begin{equation}\label{Radii}
\begin{array}{c}
\ds R_M^2\left( s\right) =R_{0M}^2+\alpha _M^{\prime }\left( 0\right) \ln \left(
-\frac s{m_0^2}\right)  \\ \\
\ds R_D^2\left( s\right) =R_{0D}^2+\alpha _D^{\prime }\left( 0\right) \ln \left(
-\frac s{m_0^2}\right)
\end{array}
\end{equation}

\noindent 
Substituting the amplitudes defined by eq.(\ref{PionNucleonFragmentation})
into factorization formula (\ref{factorization}) we get:

\begin{equation}\label{CrossAmplitude}
\begin{array}{c}
\ds \widetilde{A}^{\pi \overline{\pi }\rightarrow
N\overline{N}}(s,{\bf b}_{\perp })= \\ \\
\ds  \left( N_MN_D\right) ^{1/2}
\frac 1{4\pi R_D\left( s\right) R_M\left(  s\right)
}\left( -\frac s{m_0^2}\right) ^{\frac 12\left( \alpha _D\left( 
0\right) + \alpha _M\left( 0\right) \right) } 
\exp \left[ -{\bf b}_{\perp }^2 \left(
\frac 1{8R_M^2\left( s\right) }+\frac 1{8R_D^2\left( s\right) }\right) \right]
\end{array}
\end{equation}

For consistency of eqs. (\ref{CrossAmplitude}), (\ref{MBDReggeAmplitude}) we
should require the following relations between Regge parameters and
normalization constants \cite{Kaidalov1}:

\begin{equation}\label{OurPlanar} \begin{array}{c}
\ds 2\frac 1{R_B^2\left( s\right) }=\frac 1{R_M^2\left( s\right)
}+\frac 1{R_D^2\left( s\right) } , \\ \\
\ds 2\alpha \left( 0\right) _B=\alpha _D\left( 0\right) +\alpha _M\left(
0\right) , 
\end{array}
\end{equation}

\begin{equation}
\label{NormCnst}\left( N_MN_D\right) ^{1/2}\frac 1{R_D\left( s\right)
R_M\left( s\right) }=N_B\frac 1{R_B^2\left( s\right) } 
\end{equation}

If only light $u$, $d$ quarks are involved we can safely assume that \cite{Kaidalov1}

\begin{equation}\label{UDPlanar}
\begin{array}{c}
\ds \alpha _M^{\prime }\left( 0\right) =\alpha _B^{\prime }\left(
0\right) =\alpha _D^{\prime }\left( 0\right)\equiv  \alpha ^{\prime }\left(
0\right)\\ \\ 
\ds R_{0M}^2\left( 0\right) =R_{0B}^2\left( 0\right) =R_{0D}^2\left(
0\right)  \equiv  R_{0}^2\left( 0\right)\\ \\
\left( N_MN_D\right) ^{1/2}=N_B 
\end{array} 
\end{equation}

Then the relations (\ref{OurPlanar}), (\ref{NormCnst}) can be satisfied at all
$s$. Otherwise, they can be satisfied at sufficiently large $s$ (see also
\cite{Kaidalov1}).

Using the same approach we can consider the reaction of
$e^{+}e^{-}$-annihilation into hadrons (see Figs. 2a,2b). In the case of
$\gamma \rightarrow N\overline{N}$ reaction a virtual photon creates a
$q\overline{q}$-pair, which forms a color string in the intermediate state.
Then, the mechanism of producing a hadron pair in the final state is the same
as it was considered above. The string breaks producing diquark-antidiquark $d
\overline{d}$-pair, which combines with spectator quarks to produce an $N 
\overline{N}$ final state. The form factors of $\gamma \rightarrow
N\overline{N }$ transition can be expressed through the amplitudes $T^{\gamma
\rightarrow q \overline{q}}\left( s, {\bf b}_{\perp } \right) $,
$\widetilde{T}^{q\overline{q}\rightarrow N \overline{N}}\left( s, {\bf
b}_{\perp }\right) $ as:

\begin{equation}\label{TriangleNN}
A^{\gamma\rightarrow N\overline{N}}\left( s\right) = \frac
i{2s}\ T^{\gamma \rightarrow q\overline{q}}\left( s\right) \cdot \widetilde{
T }^{q\overline{q}\rightarrow N\overline{N}}\left( s,{\bf b}
_{\perp}=0\right)
\end{equation}
Therefore, at large $s$ we have the following behavior of nucleon form factors: 

\begin{equation}\label{FFTriangleNN}
\left| G_{m,e}\left( s\right) \right| \sim \left|
s\right| ^{-1}\left| \widetilde{T}^{q\overline{q}\rightarrow N\overline{N}
}\left( s,{\bf b} _{\perp }=0\right) \right| \sim \frac 1{R_D\left(
s\right) }\left|  s \right|^{(\alpha _D\left( 0\right) -1)/2} 
\end{equation}

In the case of  $\pi \overline{\pi}$ pair produced in the final state (Fig.2a), 
the formulas for the amplitude and form factor can be written as:

\begin{equation}\label{TrianglePiPi}
A^{\gamma \rightarrow \pi\overline{\pi}}\left( s\right)
=\frac i{2s}\ T^{\gamma \rightarrow q\overline{q}}\left( s\right) \cdot 
\widetilde{T }^{q\overline{q}\rightarrow \pi\overline{\pi}}\left( s,{\bf b}
_{\perp }=0\right) 
\end{equation}

\begin{equation}\label{FFTrianglePiPi}
\left| F_{\pi}\left( s\right) \right| \sim \left|
s\right| ^{-1}\left| \widetilde{T}^{q\overline{q}\rightarrow
\pi\overline{\pi}} \left( s,{\bf b}_{\perp }=0\right) \right| \sim \frac
1{R_M\left( s\right) }\left|  s\right| ^{(\alpha _M\left( 0\right)
-1)/2}
\end{equation}

Taking the values $\alpha_{M}(0)\simeq -0.5$, $\alpha_{M}(0)\simeq -1$,
$\alpha_{\left( D\right) }\left( 0\right) =(2\alpha _B\left( 0\right) -
\alpha_M\left( 0\right) )\simeq -1.5$ we get $$F_{\pi}(s) \sim
\left|s/s_0\right|^{-1/4},$$ $$\left| G_{m,e}\left( s\right) \right| \sim
\left| s/s_0\right| ^{-5/4}.$$ Such asymptotic of the form factors differs
from the  predictions of the quark counting rules \cite{BrodskyFarrar} and
{\em pQCD} calculations \cite{Brodsky}. Note also that this power behavior
does not  agree also with  experimental behavior for pionic $F_\pi(s) \sim
s^{-1}$ and nucleon form factors $G_{m,e}\sim s^{-2} $. The main reason of
this disagreement is that we ignored up to now additional suppression caused
by Sudakov form factor. It will be demonstrated in Sections \ref{spin} and
\ref{sudak} that taking into account the Sudakov form factor we shall be able
to describe correctly $q^2$ dependences of the pion and nucleon form factors
as well as to explain their differences in the space-like and time-like
regions. However, before coming to the discussion of Sudakov form factor  and
to numerical calculations we should at first consider  spin effects and
possibility to separate  charge and magnetic nucleon form factors. We shall
also demonstrate that the chirality conservation leads to additional
suppresion of the pion form factor $\sim 1/\sqrt{s}$ at large $s$.

\section{ Spin structure of amplitudes and form factors.}\label{spin}

Let us introduce necessary notations. The pion and nucleon form factors can
be defined as follows:  \begin{equation}
\label{PionFFDef}
A^{\gamma\rightarrow \pi 
\overline{\pi}}_{\mu}(s)=
F_{\pi}\left( p_{h\mu} - p_{\overline{h}\mu}\right)\hspace{0.2cm}
\end{equation}
and

\begin{equation}
\begin{array}{c}
\label{NucleonFFDef} 
\ds A^{\gamma\rightarrow N   \overline{N}}_{\mu}(q^2)=
\overline{u}_{\lambda_N}\left[  G_{m}(q^2)\gamma_{\mu}+2M( G_e(q^2) - G_m (q^2) )
\frac{\ds \left(p_{h} - p_{\overline{h}}\right)_\mu}
{\ds \left( p_{h} - p_{\overline{h}}\right)^2}\right]
\upsilon_{\lambda_{\overline{N}}} =\\ \\ 
 = \ds \overline{u}_{\lambda_N}\left[  
F_1(q^2)\gamma_\mu - \frac{ \kappa_p F_2 (q^2) }{2 M_h}\sigma_{\mu\nu} q_\nu 
\right] \upsilon_{\lambda_{\overline{N}}} \ ,
\end{array}
\end{equation}

\noindent 
where $p_h$ and $p_{\overline{h}}$ are the 4-momenta of final hadrons, $\kappa_p = 
\mu_p - 1$ and $\mu_p$ is the proton magnetic moment $\mu_p=2.793$.
 
Our aim is to introduce the spin structure of the amplitudes $A^{q
\overline{q} \rightarrow \pi\overline{\pi}}$ and $A^{q \overline{q}
\rightarrow N\overline{N}}$. We define the invariants $s$ and $t$ for a
two-body scattering amplitude $q\overline{q} \rightarrow h \overline{h}$ in
the standard way

\begin{eqnarray} \label{Invariants}
&&s=\left( p_{q}+p_{
\overline{q}}\right)^2= \left( p_{h}+p_{\overline{h}} \right)^2 , \nonumber\\ 
&&t=\left( p_q - p_{h} \right)^2= \left( p_{\overline{q}} - p_{\overline{h} 
}\right)^2
\end{eqnarray}

\noindent and introduce the relative momenta $ p_{\mu} = 
\frac{1}{2}(p_{q \mu }-p_{\overline{q},\mu})$, $ P_{\mu} =
\frac{1}{2}( p_{h\mu}-p_{\overline{h},\mu})$, $k_{\mu}=P_{\mu}-p_\mu$, where
$p_{q\mu}$, $k_\mu$ and $P_{h\mu}$ are the four-momenta of the quark, antiquark
(or diquark) and hadron respectively.

At large $s$ and finite $t$ we have $\{\vec{ P}
^2,\hspace{0.05cm} \vec{p}\ ^2 \} \gg \{ M^2,\hspace{0.05cm}m^2\}$, where $m$ is 
the mass of the light quark and ${\bf P}_\perp=0, P_z$. Then the following 
relation for the longitudinal momenta:  
\begin{equation}\label{kz}
k_z=P_z -p_z \approx \frac{\sqrt{s}}{2}- \frac{M^2}{\sqrt{s}} -
\left(\frac{\sqrt{s}}{2}- \frac{m^2+{\bf k}_\perp^2}{\sqrt{s}}\right) \sim 
\left(
\frac{M^2}{\sqrt{s}},\frac{m^2+{\bf k}_\perp^2}{\sqrt{s}}\right)
\end{equation}
is satisfied. As it follows from (\ref{kz}) $k_z$ can be neglected as compared 
to $\vert {\bf k}_{\perp} \vert$: $k_z \approx 0$.

Let us consider the pion form factor. The amplitude $T^{q\overline{q}\rightarrow 
\pi \overline{\pi}}$ can be expressed through two invariant amplitudes which 
correspond to the odd and even angular momenta of the $\pi^{+} \pi^{-}$ pair:

\begin{equation}
\label{PionQuarkAmplitude0}
T_{\lambda_q \lambda_{\overline{q}}}^{q\overline{q} \rightarrow \pi 
\overline{\pi}} \left( s, {\bf k}^2_{\perp}\right) = \left(
\chi^{\star}_{\lambda_{\overline{q}}} \sigma_i\chi_{\lambda_q} \right)
\left[ \widetilde{B}_1 \left( s, {\bf k}^2_{\perp}\right) P_i+ 
\widetilde{B}_2\left( s, {\bf k}^2_{\perp}\right) p_i \right] .
\end{equation}

We assume that the amplitudes $\widetilde{B}_1$, $\widetilde{B}_2$ have the
same asymptotics at large $s$ and parameterize them in the following way: 
\begin{equation}
\label{BiOddEvenAkiRegge}
\widetilde{B}_{\{1,2\}} \left( s, t \right) = \beta^{0}_{\{1,2\}} \left(- 
\frac{s}{ s_0}\right)^{\alpha_M(0)/2} \exp \left[ \frac{1}{2} \left(
R_{0}^2+\alpha^{\prime}(0)\ln\left( - \frac{s}{ s_0} \right) \right) t \right].
\end{equation}
The factor $s^{1/2}$ has been removed from $\ds \left
(\frac{s}{s_0}\right)^{(\alpha_M(0)+1)/2}$ (see eq.(8)) to account for 
parameterization
(\ref{PionQuarkAmplitude0}) where $\vert \vec{P} \vert , \vert \vec{p} \vert
\sim s^{1/2}$.

It is convenient to rewrite eq.(\ref{PionQuarkAmplitude0}) in the following 
form:

\begin{equation}
\label{PionQuarkAmplitude1}
T_{\lambda_q \lambda_{\overline{q}}}^{q\overline{q} \rightarrow \pi 
\overline{\pi}} \left( s, {\bf k}^2_{\perp}\right) = \left(
\chi^{\star}_{\lambda_{\overline{q}}} \sigma_i\chi_{\lambda_q} \right)
\left[ B_1 \left( s, {\bf k}^2_{\perp}\right) P_i+ B_2 \left( s, {\bf 
k}^2_{\perp}\right) k_i \right] .
\end{equation}
Here $B_1\left( s, {\bf k}^2_{\perp}\right)$, $B_2\left( s, {\bf k}
^2_{\perp}\right)$ have the same dependence on $s$ and ${\bf
k}^2_{\perp}$ (see (\ref{BiOddEvenAkiRegge})) as $\widetilde{B}_1\left( s,
{\bf k}^2_{\perp}\right)$, $ \widetilde{B}_2\left( s, {\bf
k}^2_{\perp}\right)$. As $\vert {\bf k}\vert /\vert \vec{P}\vert \sim
s^{-1/2}$, the contribution of the second term of (\ref{PionQuarkAmplitude1})
into form factors is parametrically small and can be neglected (see also
below).

The expression for the matrix element of the current operator $A_\mu^{\gamma
\rightarrow \pi \overline{\pi}}$ can be written as the convolution of the
quark current and quark-pion transition amplitude (\ref{FFTriangleNN})

\begin{equation}\label{CurrentPionDef}
A^{\gamma \rightarrow \pi\overline{\pi}}_{\mu}(s) = \frac{i}{(8 \pi^2 s)}
\int{\ d^2 {\bf k}_{\perp}\;  T_{\mu}^{ \gamma\rightarrow q \overline{q}}
\left( s, {\bf k}_{\perp}^{2}  \right) T^{ q \overline{q}\rightarrow \pi
\overline{\pi}} \left( s, {\bf k}_\perp^{2} \right) }\ .
\end{equation}

The quark current operator $A^{\gamma \rightarrow q \overline{q}}_{i}= 
\overline{u}(p_q,\lambda_q)\gamma_i \upsilon(p_{\overline{q}},\lambda_{
\overline{q}})$ can be written in c.m.s., where $
\vec{p}
_{q}=-\vec{p}_{\overline{q}}=\vec{p}$,  in the following form:

\begin{equation}
\label{QuarkCurrentCMS}
\begin{array}{c}
\ds
A^{\gamma \rightarrow q \overline{q}}_i = 2 (\varepsilon +m)^{-1} 
\chi_{\lambda_{q}}^{\star} \left[ \varepsilon (\varepsilon +m)\sigma_i - p_i
\vec p \cdot \vec{\sigma}) \right] \chi_{\lambda_{\overline{q}}} \ .
\end{array}
\end{equation}

\noindent For the following applications it is convenient to separate the
quark  current operator into transversal and longitudinal parts:

\begin{equation}\label{QuarkCur2}
\begin{array}{c} \ds A^{\gamma \rightarrow q \overline{q}}_i =  2
\left[ \varepsilon \left( \delta_{uj}-\frac{p_i p_j}{\vec{p}\ ^2} \right)+  m
\ds \frac{p_ip_j}{\vec{p}\ ^2}\right]
(\chi_{\lambda_{q}}^{\star}\sigma_j\chi_{\lambda_{\overline{q}}}).
\end{array}
\end{equation}

\noindent Taking into account spin variables in (\ref{CurrentPionDef}) we can 
express the transition amplitude $\ds A^{\gamma \rightarrow \pi \pi}$ in the
following form

\begin{equation} \label{CurrentPion1}
\begin{array}{c}
\ds A^{\gamma \rightarrow \pi  \pi}_{i}(s) =  \\ \\
\ds \frac{i}{(8 \pi^2
s)} \int d^2 {\bf k}_{\perp} \ 2 Tr \left[\sigma_j\sigma_l\right] 
\left[ \left( \varepsilon -(\varepsilon -m)\frac{{\bf
k}^2_\perp}{2\vec{p}\ ^2}\right) \delta_{ij}^\perp +
\left( m +(\varepsilon -m)\frac{{\bf
k}^2_\perp}{\vec{p}\ ^2}\right)\frac{P_iP_j}{\vec{P}\ ^2}
\right]\times \\ \\
\ds \times \left[ B_1 \left( s,{\bf k}^2_{\perp}\right)
P_l+ B_2 \left( s, {\bf k}^2_{\perp}\right) k_l \right] ,
\end{array}
\end{equation}
where $\ds \delta^{\perp}_{ij} = \delta_{ij} -\frac{P_i P_j}{\vec{P}\ ^2}$.
It is easy to see that the term proportional to $B_2 \left( s, {\bf
k}^2_{\perp} \right)$ gives relative contribution $\ds \sim \frac{m^2 +{\bf
k}_\perp^2 }{ s }$ as compared with the term proportional to $B_1 \left( s,
{\bf k}^2_{\perp} \right)$. Therefore we get:

\begin{equation} \label{CurrentPion2}
\begin{array}{l}
\ds A^{\gamma \rightarrow \pi  \overline{\pi}}_{i}(s) = 
\frac{i}{(8 \pi^2 s)}  \int d^2 {\bf k}_{\perp}\ 2 
\left[ m +(\varepsilon -m)\frac{{\bf k}^2_\perp}{\vec{p}\ ^2}\right]
 B_1 \left( s,{\bf k}^2_{\perp}\right)  P_i \approx \\ \\ 
\ds \approx  \frac{i}{(8 \pi^2 s)}  \int d^2 {\bf k}_{\perp} \ 2 m \ B_1
\left( s, {\bf k}^2_{\perp}\right) P_i \ ,
\end{array}
\end{equation} 
where in the last expression all the pre-asymptotic terms of the order of 
$s^{-1/2}$,  $ s^{-1}$ and so on are neglected.

It is important to note that the transversal and longitudinal components of
the quark curent $\gamma \rightarrow q\overline{q}$ behave differently at
large $s$:

\begin{equation}  \label{AzAperpPion} 
\begin{array}{c}
\ds {\bf A}_\perp^{\gamma \rightarrow q\overline{q}} \sim \varepsilon {\bf
\sigma}_\perp \sim s^{1/2} {\bf \sigma}_\perp , \\ \\
\ds A_z^{\gamma \rightarrow q\overline{q}}\sim m \sigma_z \ .
\end{array}
\end{equation}
The small value of the longitudinal component is related to the conservation of
chirality. This component vanishes in the limit $m\rightarrow 0$.

In the case of the transition $\gamma \rightarrow \pi^+ \pi^-$ only the 
longitudinal components of the quark current gives contribution, which is 
suppressed at large $s$ as $\ds \sim s^{-1/2}$. That is the reason why the pion 
form factor has additional suppression $\ds \sim
s^{-1/2}$ as compared with estimation (\ref{FFTrianglePiPi}), which was 
presented in the previous section, where spin effects were neglected. 
Note that such behaviour of the pion form factor was predicted by Ravndal 
\cite{Ravndal} in parton model. Such a suppression should be absent in the 
transitions $\gamma \rightarrow \pi \rho$ and è $\gamma \rightarrow \pi \omega$.
Therefore QGSM predicts that the ratios of the form factors $\ds
\frac{F_{\pi\pi}}{F_{\pi \rho}}$ and $\ds \frac{F_{\pi\pi}}{F_{\pi \omega}}$
should decrease with $s$ as $\sim \ds \frac{1}{\sqrt{ s }}$. Note that in the 
perturbative QCD the quark helicities are conserved, and this would lead to the 
different behaviour of those ratios $\ds \left(
\frac{F_{\pi\pi}}{F_{\pi\rho}} \right)_{PQCD} \sim \left(
\frac{F_{\pi\pi}}{F_{\pi \omega}}\right)_{PQCD} \sim s $. Therefore the QGSM 
predictions for the form factor ratios $\gamma \rightarrow \pi \pi$ and è $\gamma
\rightarrow \pi \rho (\omega)$ at large $ s $ are very different from the 
PQCD predictions. Note that in the nucleon case both components $A_{\perp}$ and è 
$A_z$ contribute and there is no such power suppression of the nucleon 
charge or magnetic form factor (see eq.(\ref{CurrentNucleon3})).

Let us show that from two invariant amplitudes $B_1( s,{\bf k}_{\perp  }^2)$
and $B_2( s,{\bf k}_{\perp  }^2)$, defined by eq.(24), the main contribution
into the planar pion diagram of Fig.1a) is given by the first amplitude $B_1(
s,{\bf k}_{\perp  }^2)$:

\begin{eqnarray} \label{4PionWithSpin} 
&&A^{\pi \overline{\pi }\rightarrow \pi \overline{\pi }}( s,{\bf
q}_\perp^2)  =\frac i{(8\pi ^2s)}\sum_{\lambda _q\lambda _{\overline{q}}}\int
d^2{\bf k}  _{\perp }
( \chi _{\lambda _q}^{\star }\sigma _i\chi_{\lambda_{\overline{q}}})
( \chi_{\lambda _{\overline{q}}}^{\star }\sigma_j\chi _{\lambda _q})
\times   \nonumber \\  
&&\times \left[ B_1^{\star }( s,{\bf k}_{\perp }^2) 
P_i+B_2^{\star }( s,{\bf k}_{\perp }^2) k_i\right]
\left[ B_1( s,({\bf q-k})_{\perp }^2) P_j+
B_2( s,({\bf q-k})_{\perp  }^2) k_j\right] = \\  
&& =\frac i{(8\pi ^2s)}\int d^2{\bf k}_{\perp }\left[ B_1^{\star }( s,{\bf
k}_{\perp }^2)P_i+B_2^{\star }( s,{\bf k}_{\perp}^2)
k_i\right]  \times \left[ B_1( s,( {\bf q-k})_{\perp }^2) P_i+B_2(
s,({\bf q-k})_{\perp }^2)  k_i \right] =\nonumber \\   
&&=\frac{i}{(8\pi ^2s)}\int d^2{\bf k}_{\perp
}B_1^{\star }( s,{\bf k}_{\perp }^2) B_1( s,({\bf q-k})_{\perp
}^2) s+O \left( \frac{{\bf k}_\perp^2}{s},\ \frac{m_q^2}{s}\right) . \nonumber 
\end{eqnarray}

Provided that $ B_1\sim \left(-\ds \frac s{s_0}\right)
^{\alpha_M(0)/2}$  eq.(\ref{4PionWithSpin}) garantees
the correct $s$-dependence of \\ $A^{\pi \overline{\pi }\rightarrow \pi
\overline{\pi }}\sim \left( -\ds \frac{s}{s_0}
\right)^{\alpha_M(0)}$. Moreover, eq.(\ref{4PionWithSpin})  can be used
to fix the normalization constant for $B_1\left( s, (  {\bf
q-k})^2_{\perp} \right)$.

The pion form factor $F_{\pi}$ can be expressed directly through $B_1 \left(
s, ( {\bf q-k})^2_{\perp} \right)$:

\begin{equation}\label{PionFFExpr}
F_{\pi}(s) \simeq m \frac i{(8\pi ^2s)} \int
d^2 {\bf k}_{\perp} B_{1}\left( s, ({\bf q-k})^2_{\perp} \right) = N_\pi
\frac{1}{\sqrt{\left(R_0^2+\alpha^{\prime}\ln
\left( -\ds \frac{s}{s_0} \right)\right)}} \left( -
\frac{s}{s_0}\right)^{(\alpha_{M0}/2-1)} \ ,
\end{equation} 
where $N_\pi$ is a normalization constant, which we consider here as a free 
parameter.

Let us now discuss nucleon form factors. The $q\overline{q} \rightarrow N
\overline{N}$ transition amplitude can be parameterized in terms of eight
invariant amplitudes:

\begin{equation}\label{NucleonQuarkAmplitude}
\begin{array}{c}
T^{q\overline{q} \rightarrow N \overline{N}} \left( s,{\bf k}
_{\perp}\right) =
\chi^{\star}_{\lambda_{\overline{N}}} \chi^{\star}_{\lambda_N}
U_{( \lambda_{\overline{N}} \lambda_N \lambda_q  \lambda_{\overline{q}})} 
\chi_{\lambda_q} \chi_{\lambda_{\overline{q}}} \ ,
\end{array}
\end{equation}
where $U$ is a spin-spin operator that acts on spin variables of nucleons and
quarks. The general spin structure of $U$ can be written in terms
of eight invariant amplitudes: 
\begin{eqnarray}\label{U} 
U&=&D_1 \left( s, {\bf q}_{\perp} \right) 1\cdot 1 + 
D_2\left( s, {\bf k}_{\perp} \right) (\vec{\sigma}\; \vec{n})\cdot 1 +
D_3\left( s, {\bf k}_{\perp} \right) 1\cdot (\vec{\sigma}^{\prime}\; \vec n)+
D_4\left( s, {\bf k}_{\perp} \right) \sigma_{x}\cdot \sigma^{\prime}_{x}+
\nonumber \\
&+&D_5\left( s, {\bf k}_{\perp} \right) \sigma_{y} \cdot\sigma^{\prime}_{y}+ 
D_6\left( s, {\bf k}_{\perp} \right) \sigma_{z} \cdot\sigma^{\prime}_{z}+ 
D_7\left( s, {\bf k}_{\perp} \right) \sigma_{x} \cdot\sigma^{\prime}_{z}+
D_8\left( s, {\bf k}_{\perp} \right) \sigma_{z} \cdot \sigma^{\prime}_{x} 
\end{eqnarray} 
Here matrices $\sigma_i$ and $\sigma_i^{\prime}$ act on spin indices of
nucleon and quark (antinucleon and antiquark) respectively, and $\vec{n}$ is the
unit vector normal to the scattering plane:

\begin{equation} \label{UnitaryVectorsDefined} 
\vec n=\frac{[\vec P\times \vec p \hspace{0.1cm}]} {\left|[ \vec P\times \vec p
\hspace{0.1cm}]\right|}\ .
\end{equation}

At first we assume that all the amplitudes $D_i$ have the same Regge asymptotic: 
\begin{equation}\label{DiAkiRegge}
D_{i} \left( s, t \right) =
\gamma^{0}_{i} \left(- \frac{s}{ 
s_0}\right)^{(\alpha_D(0)+1)/2} \exp \left[ \frac{1}{2}\left(
R_{0}^2+\alpha^{\prime}(0)\ln\left( - \frac{s}{  s_0} \right) \right)
t \right] \ .
\end{equation}
Then the amplitude of the transition $ \gamma \rightarrow
N\overline{N}$ can be written as:

\begin{equation}\label{CurrentNucleon2}
\begin{array}{c}
\ds A^{\gamma \rightarrow N
\overline{N}}_{i}(s) = \frac{i}{(8 \pi^2 s)} \int{ d^2 {\bf k}_{\perp}\; 
T_{i}^{ \gamma\rightarrow q \overline{q}} ( s, {\bf k}_{\perp} 
) \hspace{0.2cm} T^{ q \overline{q}\rightarrow N \overline{N}}
( s, {\bf k}_\perp ) } = \\   \\ 
\ds = \frac{i}{(8 \pi^2 s)} \sum_{\lambda_q \lambda_{\overline{q}}}\int d^2
{\bf k}_{\perp} \ 
2 \left[  
\left( \varepsilon -(\varepsilon -m)\frac{{\bf k}^2_\perp}{2\vec{p}\ ^2}
\right) \delta_{ij}^\perp + \left( m +(\varepsilon -m)\frac{{\bf
k}^2_\perp}{\vec{p}\ ^2}\right)\frac{P_i P_j}{\vec{P}^2} \right]
\times \\ \\
\times \left( \chi_{\lambda_q}^{\star}
\sigma_j  \chi_{\lambda_{\overline{q}}}\right) 
\chi^{\star}_{\lambda_N}\chi^{(c)\star}_{ \lambda_{\overline{q}}}
\hat{U}_{( \lambda_N \lambda_{\overline{N}} ;\ \lambda_q 
\lambda_{\overline{q}})} \chi_{\lambda_q}  \chi^{(c)}_{\lambda_{\overline{N}}}
= 2 \left[ \ds \varepsilon  G_{m} \delta^\perp_{ij} + M  G_e \ds
\frac{P_i P_j}{\vec{P}^2} \right] (\chi^{\star}_{\lambda_{N}} \sigma_j 
\chi^{(c)}_{\lambda_{\overline{N}}}) .
\end{array} 
\end{equation}

Let us write now expressions for the longitudinal and transversal components of 
the current matrix elements. It is clear from eq. (\ref{CurrentNucleon2}) that 
Sachs form factors can now easily be separated: $G_e$ contributes to 
longitudinal component only and $G_m$ to the transversal one. It is also
clear that the main asymptotic terms in quark current are not  dependent on
${\bf k}_\perp$. Taking into account that after integration over  $d^2 \ {\bf
k}_\perp$ the terms which are not invariant under rotation around  $z$-axis
should disappear, we can write the longitudinal and transversal  components of
current operator in the following form:

\begin{eqnarray}\label{CurrentNucleon3}
\ds A^{\gamma \rightarrow N\overline{N}}_z= \ds  \frac{i}{(8 \pi^2 s)} \times
\hspace{12cm} \nonumber \\ 
\times \sum_{\lambda_q
\lambda_{\overline{q}}}
\sum \limits_{l}
\int d^2 {\bf k}_{\perp}\hspace{0.1cm} D_l(s, {\bf k}^2_{\perp}) 2m   
\chi^{\star}_{\lambda_N} \sigma_{n_l} \chi^{(c)\star}_{\lambda_{\overline{q}}} 
\left(\chi_{\lambda_q}^{\star}\sigma_z\chi_{\lambda_{\overline{q}}}\right)
\chi_{\lambda_q} \sigma_{n_l} 
\chi^{c}_{\lambda_{\overline{N}}} = 2M\hspace{0.1cm}G_e (\chi^{\star}_{\lambda_N} \sigma_z
\chi^{(c)}_{\lambda_{\overline{N}}} ) ,  \nonumber \\ 
\ds  {\bf A}^{\gamma \rightarrow N\overline{N}}_{\perp}=  \frac{i}{(8 \pi^2
s)} \times \hspace{12cm} \\
\ds \times 
\sum_{\lambda_q \lambda_{\overline{q}}} \sum \limits_{l}
\int d^2 {\bf k}_{\perp}\hspace{0.1cm} D_l(s, {\bf k}^2_{\perp})\
(2\varepsilon) \chi^{\star}_{\lambda_N} \sigma_{n_l} \chi^{(c)\star}_{\lambda_{\overline{q}}} 
\left(\chi_{\lambda_q}^{\star}{\bf
\sigma}_\perp\chi_{\lambda_{\overline{q}}}\right) \chi_{\lambda_q}
\sigma_{n_l}  \chi^{c}_{\lambda_{\overline{N}}}= 
2\varepsilon\hspace{0.1cm}G_m (\chi^{\star}_{\lambda_N}
{\bf \sigma}_{\perp} \chi^{(c)}_{\lambda_{\overline{N}}} ), \nonumber
\end{eqnarray}
where  $l = 1,4,5$ and $6$. It means that the form factors $G_e$ and $G_m$ 
can be expressed through linear superpositions of the amplitudes 
$D_1$, $D_4$, $D_5$ and $D_6$. If all these amplitudes will have the same 
asymptotics then the asymptotics of $G_e$ and $G_m$ will also be the same:

\begin{equation}\label{GmGeAsymptotics}
\begin{array}{c}
\ds G_m(s)\ , \ G_e(s) \ \sim \left(-
\frac{s}{s_0}\right)^{(\alpha_D(0)-1)/2}\ .
\end{array}
\end{equation}
However the ratio $\ds \frac{G_e}{G_m}$ can be dependent on the model of 
the invariant amplitudes.

It follows also from eqs.(\ref{CurrentNucleon3}) that the Pauli form factor 
$F_2$ is suppressed at large $s$ as $s^{-1}$ with respect to $G_m$ and 
$G_e$. This suppression is purely kinematical and follows from the definition
of $F_2$:

\begin{equation}\label{F2ThroughGeGm}
\begin{array}{c}
\ds F_2(s)=\left(  \frac{q^2}{4M^2}-1\right)^{-1} \left[ G_e(s)
-G_m(s) \right] , \hspace{1cm} \ds  F_2 (s) \sim
\left(-\frac{s}{s_0}\right)^{(\alpha_{D}(0)/2-1.5)} .
\end{array}
\end{equation}

As explicit relations between the amplitudes $ D_1$, $D_4$, $D_5$ and
$D_6$ are not known we perform calculations for two different simple models:\\ 
\noindent i) dominant contribution comes from the non-spin flip 
amplitude $D_1$;\\
\noindent ii) the amplitude $T^{q\overline{q}\rightarrow N\overline{N}}$ can be 
described by a scalar diquark exchange in the $t$-channel.

\noindent In the case i) we have:

\begin{equation} \label{IfOnlyOne} 
T^{q\overline{q}\rightarrow N \overline{N}}_{\lambda_{N}
\lambda_{\overline{N}} \lambda_{q} \lambda_{\overline{q}}} \rightarrow
D_1 \left( s, {\bf k}_{\perp}\right)1\cdot 1 \equiv
D_1 \left( s, {\bf k}_{\perp}\right) \delta_{\lambda_{N} \lambda_{q}} 
\cdot \delta_{ \lambda_{\overline{N}}\lambda_{\overline{q}} }
\end{equation} 
and the expression for the amplitude $A^{\gamma \rightarrow N \overline{N}}_i$
is essentially simplified:

\begin{equation} \label{F1F2ifOneAmplitude} 
\begin{array}{c}
\ds A^{\gamma \rightarrow N   \overline{N}}_i(s) =  \frac{i}{(8
\pi^2 s)} \int d^2 {\bf k}_{\perp}\  D_1 \left( s, {\bf k}_{\perp} \right) 
\left( \chi^{\star}_{\lambda_N} \sigma_j \chi_{\lambda_{\overline{N}}} \right)
\times \\ \\
\ds \times  2 \left[  \left( \varepsilon -(\varepsilon -m)\frac{{\bf
k}_\perp^2}{2\vec{p}\ ^2}\right) \delta_{ij}^\perp+  \left( m + (\varepsilon
-m) \frac{{\bf k}_\perp^2}{\vec{p}\ ^2} \right)\frac{P_{i} P_j}{\vec{P}^2}
\right].
\end{array} 
\end{equation}
In the leading order in $s$ we can neglect all the terms proportional to $\ds 
\frac{{\bf k}_\perp^2}{\vec{p}\ ^2}$ and write the amplitude $A^{\gamma
\rightarrow N \overline{N}}_i$ in the form:

\begin{equation} \label{F1F2ifOneReduced} 
\begin{array}{c}
\ds A^{\gamma \rightarrow N   \overline{N}}_i(s) =  \frac{i}{(8
\pi^2 s)} \int d^2 {\bf k}_{\perp}\  D_1 \left( s, {\bf k}_{\perp} \right) 
\left( \chi^{\star}_{\lambda_N} \sigma_j \chi_{\lambda_{\overline{N}}} \right)
2 \left[  \varepsilon  \delta_{ij}^\perp+   m \frac{P_{i} P_j}{\vec{P}^2}
\right]\ .
\end{array} 
\end{equation}

\noindent Then we get the following expressions for the Sachs form factors: 
\begin{equation} \label{GmGeTwoSpinors} 
\begin{array}{c}
\ds G_m(s)=\frac{i}{(8 \pi^2 s)} 
\int d^2 {\bf k}_{\perp} D_1
\left(s, {\bf k}_{\perp}\right) =\frac{C}{\sqrt{
\ds R_0^2+\alpha ^{\prime }\left( 0\right)  \ln \left( -\frac
s{s_0}\right)} } \left( -\frac{s}{s_0}\right) ^{\left( \alpha
_D\left( 0\right) -1\right) /2}\ ,\\ \\ 
\ds G_e(s)=\frac{m}{M}\frac{i}{(8 \pi^2 s)}
\int d^2 {\bf k}_{\perp} D_1 \left( s, {\bf
k}_{\perp}\right) = \frac{ \frac{m}{M} \ \ C}{\sqrt{
\ds R_0^2+\alpha ^{\prime }\left( 0\right)  \ln \left( -\frac
s{s_0}\right)} } \left( -\frac{s}{s_0}\right) ^{\left( \alpha
_D\left( 0\right) -1\right) /2}\ ,
\end{array}
\end{equation} 
where $C$ is the normalization constant. The Pauli form factor $F_2$ can be 
expressed through $G_m$, $G_e$ as follows (see eq.(\ref{F2ThroughGeGm})):

\begin{equation} \label{PauliFF} 
\begin{array}{c}
\ds F_2(s)= -\left(\frac{s}{4M^2}-1\right)^{-1} \frac{C\left(1-\ds
\frac{m}{M}\right) }{ \sqrt{ \ds R_0^2+\alpha ^{\prime
}\left( 0\right)  \ln \left( \frac s{s_0}\right)} }\cdot \left(
-\frac{s}{s_0}\right) ^{\left( \alpha _D\left( 0\right) -3\right) /2} \ .
\end{array}
\end{equation}

As it follows from eqs.(\ref{NucleonQuarkAmplitude}), (\ref{U}) the 
parametrization (36) corresponds to the proper Regge behaviour of the amplitude 
$T^{N\overline{N} \rightarrow N\overline{N}}$. Indeed the amplitude  
$T^{N\overline{N}\rightarrow N\overline{N}}$ can be expressed through the 
quadratic form

\begin{equation} \label{FourNucleonVetex} 
\begin{array}{c}
\ds T^{N\overline{N} \rightarrow N\overline{N}}(s,{\bf q}_{\perp}) = 
\frac{i}{(8 \pi^2 s)}\int d^2 {\bf k}_{\perp} D_1(s,{\bf
k}_{\perp}) D_1(s,{\bf q}_{\perp}-{\bf k}_{\perp}) \ ,
\end{array}
\end{equation} 
and therefore has the following asymptotics

\begin{equation} \label{FourNucleonRestore} 
\begin{array}{c}
\ds T^{N\overline{N} \rightarrow N\overline{N}}(s,{\bf q}_{\perp}) \sim 
\left(
-\frac{s}{s_0}\right) ^{ \alpha _D \left( - {\bf q}_\perp^2 \right)}
\ ,
\end{array}
\end{equation}
which coincides  with eq.(\ref{MBDReggeAmplitude}).

In the case ii) the transition amplitude $q\overline{q}\rightarrow
N\overline{N}$ can be expressed in the covariant way through the Dirac spinors 
of quarks and nucleons: 
\begin{equation} 
\label{AmplitudeSpinStructure} 
T^{q\overline{q}\rightarrow N\overline{N}}(\lambda_{q}\lambda_{\overline{q} 
}\lambda_{N}\lambda_{\overline{N}})= ( \overline{u}_N^{\lambda _N}
u_q^{\lambda _q} ) \cdot ( \overline{u}_{\overline{N}}^{\lambda
_{\overline{N}}} u_{\overline{q}}^{\lambda _{\overline{q}}} ) .
\end{equation}
Then the nucleon current $  A^{\gamma \rightarrow N\overline{N}}$ can be written as 

\begin{equation} \label{GmANDGe} 
\begin{array}{c}
\ds A^{\gamma \rightarrow N\overline{N}} =
\int d^2{\bf k}_{\perp}\; A\left( s,-{\bf k }_{\perp }^2\right) 
\left( \left[ \left( M+m\right) ^2+{\bf k}_{\perp }^2\right] \gamma _\mu
-2\left( M+m\right) k_\mu +2k_\mu k_\nu \gamma _\nu  \right) = \\ \\
\ds =G_m\gamma _\mu +2M\left( G_e-G_m\right) \left( \frac{P_\mu
}{P^2}\right) .
\end{array}
\end{equation}

In the first approximation we can change ${\bf k}_{\perp}^2$
to the efective value $\left\langle {\bf k}_{\perp }^2\right\rangle$, taken from 
the diquark momentum distribution in the nucleon wave function. Then we get the 
following expressions for the Sachs form factors: 
\begin{equation}\label{GmGeTilda}
\begin{array}{l}
\ds G_m= \widetilde{C} \frac{\left( M+m\right) ^2}{\sqrt{
\ds R_0^2+\alpha ^{\prime }\left( 0\right)  \ln \left( -\frac
s{s_0}\right)} }\cdot \left( -\frac{s}{s_0}\right) ^{\left( \alpha
_D\left( 0\right) -1\right) /2}, \\ \\
\ds G_e-G_m=\widetilde{C} \frac{\ds \left(1+\frac{m}{M}\right) \left(
m^2-M^2+\left< {\bf k}^2_\perp\right>\right) +\left< {\bf
k}^2_\perp\right>}{\sqrt{\ds  R_0^2+\alpha ^{\prime }\left( 0\right)  \ln
\left( -\frac s{s_0}\right)} } \cdot \left(- \frac{s}{s_0} \right)  ^{\left(
\alpha _D\left( 0\right) -1\right) /2} ,
\end{array}
\end{equation} 
where $\widetilde{C}$ is the normalization constant which in principle may be 
differenet from $C$ in eqs.(\ref{GmGeTwoSpinors}).

Note that in the limit $\left< {\bf k}^2_\perp\right>=0$
eqs.(\ref{GmGeTilda}) for $G_m$ and $G_e$ coincide with eqs.
(\ref{GmGeTwoSpinors}) and the ratio $\ds \frac{G_e}{G_m}$ is equal to the
ratio of the quark and nucleon masses:

\begin{equation}\label{123}
\begin{array}{c} \ds \frac{G_e}{G_m} = \frac{m}{M}\ .
\end{array}
\end{equation}

The pion and nucleon form factors (\ref{PionFFExpr}), (\ref{GmGeTwoSpinors}) and
(\ref{GmGeTilda}) were calculated at positive $s$. However, they can
analytically be continued to negative $s$ (and complex $s$). Therefore they
are defined in the whole $s$-complex plane.

\section{SUDAKOV FORM FACTOR.} \label{sudak}

A collinear configuration of quarks, which leads to the production of a 
two-body hadronic final state, can be formed only if no hard gluon is emitted in  
the initial state. If this condition does not hold and in the initial stage of  
interaction a hard ($\left| {\bf k}_{\perp}\right| > R^{-1}$) gluon is  
emitted, the collinearity of the $q\overline{q}$ configuration is not  
preserved and this hard gluon  develops a jet in the final state  
in addition to the $h\overline{h}$-system. The necessity to preserve the  
collinear $q\overline{q}$ configuration results in the Sudakov suppression.    
Sudakov form factor is related to the initial stage of interaction when a
high-energy   virtual photon produces a pair of fast quark and antiquark and
an intermediate   quark gluon string is not yet formed.  
  
In the double logarithmic approximation (DLA) the Sudakov form factor can be
written   as follows (for possible parameterizations of Sudakov form factor and
its role in jet production see e.g. review \cite{Dokshitser} and references
therein):

\begin{equation}  \label{Sudakov}
S^{(0)}(q^2)=C_1 \exp \left( -\frac{\alpha _S^{eff}}{2\pi
}C_F\ln   ^2\left( -\frac{q^2}{\omega^2}\right) \right) \ ,
\end{equation}  
where $\ds C_F=\frac{4}{3}$, and $\omega$ has the meaning of the
characteristic energy or transverse momentum of emitted gluons when the 
collinearity of $q\overline{q}$  configuration is still preserved. In formula
(\ref{Sudakov}) both values $\alpha _s^{eff}$ and   $\omega$ are introduced
as free  phenomenological parameters. We shall vary them within the limits
which are in   reasonable agreement with available theoretical considerations
and experimental  data.

It is important to note that the modulus of Sudakov form factor is different  
for positive and negative $s$ due to the double logarithmic term in the  
exponent:   

\begin{equation}   \label{SudakovRatio}
r_{ts}=\frac{\left| S^{(0)}\left( s\right) \right| }{  
\left| S^{(0)}\left( -s\right) \right| }\simeq \exp \left( C_F\ \frac{\alpha  
_s^{eff}} 2\pi \right)   
\end{equation}  
and for $\alpha _s^{eff}\simeq 0.4$ the ratio $r_{ts}\simeq 2$.  
  
In DLA the value of $\alpha_S$ is considered as a constant. This approximation  
is reasonable at not very high $s$, when characteristic transverse momenta of  
emitted gluons are about $\left\langle {\bf k}_{\perp }^2\right\rangle \sim $ 1  
GeV$^2$ ($\left\langle {\bf k}_{\perp }^2\right\rangle \ll s$). In this region  
$\alpha_S({\bf k}^2_\perp)$ can be considered as a constant $\alpha 
_s^{eff} \approx 0.4\div 0.5$ \cite{Simonov}-\cite{Dokshitzer2}. 
 
However, at very large $s$ it is necessary to take into account logarithmic  
dependence of $\alpha_S$ on $s$. In this case it is convenient to write 
the Sudakov form factor in the following form 
 
\begin{eqnarray} \label {SInt}  
S^{(1)}(s) \sim \exp \left[ - \frac{C_f}{2\pi} \ln\left( - 
\frac{s}{\mu_{1}^2}\right)\int_{\mu^2_{2} }^{s}  
\ds  \frac{d\mu^2}{\mu^2} \alpha_S( \mu^2) \right]   
\end{eqnarray} 
where the first logarithm $\ds \ln \frac{s}{\mu_1^2}$ corresponds to the  
integration on the longitudinal momentum (energy), and the second one $\ds \ln  
\frac{s}{\mu_2^2}$ - to the integration on the transverse momentum. Using the  
one-loop expression for $\alpha_S(\mu^2)$: 
 
\begin{eqnarray}\label{AlphaOneLoop} 
&&\alpha_S(\mu^2)=\frac{4 \pi}{\beta_0 
\ln \left( -\ds \frac{ \mu^2 }{\Lambda^2} \right) }\ , \\
\textrm{where}  &&\beta_0 = \frac{11}{3}N_c -\frac{2}{3} N_f =9\ , 
\end{eqnarray} 
we get the following expression for the Sudakov form factor  
\cite{Sterman:1977wj} 
 
\begin{equation}  \label{OneLoopSudakov} 
S^{(1)}(s) =C_2 \exp \left[ -\frac{2 C_F}  
{\beta_0} \ \ln \left( -\frac{s}{\mu^2_{1}}\right) \ln \left(  
\frac{\alpha_S(\mu_{2}^2)}{\alpha_S (s) }\right)   \right] 
\end{equation}
The choice of the sign before $s$ is defined by analytical properties of the  
form factors. The constants $\mu_1$, $\mu_2$ in(\ref{OneLoopSudakov}) will be  
considered as free parameters. 
 
The (time-like)-to-(space-like) ratio for $S^{(1)}(s)$ is in general smaller as  
compared to that of $S^{(0)}(q^2)$ from (\ref{Sudakov}). Nevertheless it can
also   reach the value $\sim$ 2 at $5\div15$ $GeV^2$.  At very large $s$ the
ratio $r_{ts}^{(1)}$ can be written as  

\begin{equation}  
\label{SudakovRatioOneLoop}r_{ts}^{(1)}\vert_{s \to \infty} =\exp  
\left[ \frac{2 C_F} {\beta_0} \frac{\pi^2}{\ln (s)} \right]   
\end{equation}  
(compare with eq.(\ref{SudakovRatio})) and slowly decreases with $s$. At very  
large $s$ in (\ref{SudakovRatioOneLoop}) 
the ratio $r_{ts}^{(1)}\rightarrow 1$ and Sudakov suppression disappears. 
 
Effective power which characterizes decrease of hadronic form factors with  
account of the Sudakov form factor (\ref{OneLoopSudakov}), can be defined as \\ 
$\ds   G \sim  \left( \frac s{s_0}\right) ^{\left(\alpha 
_{D}(0)-1\right)/2}  S^{(1)}(s) \sim \left( -\frac s{s_0}\right) ^{\xi_{N}
\left(  s\right) }$, $ $ $\ds   F_{\pi} \sim  \left( \frac s{s_0}\right) 
^{\left(\alpha _{M}(0)-2\right)/2}   S^{(1)}(s) \sim \left( 
-\frac s{s_0}\right) ^{\xi_{\pi} \left( s\right) }$, where \\ $\ds \xi_{N} 
\left(  s\right) =-\frac{2 C_f}{\beta_0 } \ln \left[ 
\ln{ (- s)} \right]  +\frac{\alpha_{D}( 0) -1 }{2}$  for nucleon form factors,  
and \\ 
$\ds \xi_\pi \left(  s\right) =-\frac{2 C_F}{\beta_0 } 
\ln \left[ \ln{( -s) } \right] + 
\frac{\alpha_{M}( 0) - 2}{2}$ for pion form factor.

Therefore hadronic form factors, calculated in QGSM, decrease faster than any 
Finite power of $s$. This means that at very large $s$ the perturbative  
contribution, which is proportional to $\ds \sim \frac{\alpha_s(s)}{s}$ and 
$\ds \sim \frac{\alpha^2_s(s)}{s^2}$ for pion and nucleon form factors  
respectively, becomes dominant. However, the decrease of $\xi_{N,\pi}(s)$
with $s$, which is defined by the  term  $\ds \ln{[\ln{(-s)}]}$, is rather
slow and the scale $\widetilde{s}$, where  nonperturbative contribution
becomes comparable with perturbative one, is much   larger than experimentally
available values of $s$ ($\widetilde{s}>10^2$ $GeV^2$).

Moreover we should stress that at available now values $s \simeq  30 \div 50$
GeV$^2$ the nonperturbative contribution is dominant  and the gluon virtuality
is not large ($\mu^2 \sim 1$ $GeV^2 \ll s$). The value  of $\alpha_S(\mu^2)$
in this region  of $\mu ^2$ is frozen and practically does not depend on
$\mu^2$ \cite{Simonov}.

Thus the $s$ dependence of hadronic form factors in QGSM has very transparent  
interpretation: the probability for a virtual photon to produce a two-body  
hadronic state contains two suppression factors. The first one is related to
a  suppression of the production of a collinear  $q\overline{q}$ pair. It can
be calculated using the perturbative QCD and is described by the Sudakov
form factor. The second small factor is related to  suppression of the
production of a two-body hadronic state after hadronization of a
$q\overline{q}$ string at large $s$. This factor is related to the large  
distance physics and has nonperturbative nature.

\section{Numerical Results. Comparison with Experiment.}\label{exper}

In figs. $3 - 8$  we compare the results of our calculations for hadronic
form   factors $G_{m}\left(q^2\right)$, $G_e\left(q^2\right)$,
$F_2\left(q^2\right)$   and  $F_\pi\left(q^2\right)$ with experimental data in
the time-like  and space-like regions of $q^2$.  Parameters $m$, $\left< {\bf
k}_\perp^2\right>$ and $R_0^2$ were taken to be   equal to $m=0.22$ $GeV$,
$\left< {\bf k}_\perp^2\right> =0.2$ $GeV^2$ and  $R_0^2 = 3$ $GeV^2$.

Both parameterizations of Sudakov form factors were employed:\\   
$a)$ $S^{(0)}(q^2)$ (see (\ref{Sudakov})) with frozen $\alpha_S^{eff}$; \\ 
$b)$ $S^{(1)}(q^2)$ (\ref{OneLoopSudakov}).

The best fit of nucleon form factors in the case $a)$ was found at:  
$\alpha_s^{eff}=0.45$, $\omega^2=0.35$ $GeV^2$. In the case $b)$ the
following values of parameters were chosen: $\Lambda_{QCD}=0.4$ $GeV$,
$\mu_1=1.3$ $GeV$ and $\mu_2=0.6$ $GeV$.

In Fig. 3 we present the proton magnetic form factor in the space-like region of  
$Q^2=-q^2=-s$. The bold and thin solid curves are calculated for the model 
$a)$ and $b)$ of Sudakov form factor respectively. The normalization constant 
$C$ was also considered as a free parameter in each case. Both curves describe 
the $Q^2$ dependence of $G_m$ rather well. The bold and thin dashed curves 
are calculated dividing the corresponding solid curves by Sudakov form factor. 
It is clear that the effect of Sudakov form factor is quite important: at  
$Q^2=5$ $GeV^2$ the dashed and solid curves are different by approximately 
factors of $2$ and  $1.5$ for models $a)$ and $b)$ respectively.

In Fig. 4 we present the proton magnetic form factor in the time-like region. 
The meaning of the curves is the same as in Fig. 3. The normalization factor 
in the time-like region was taken the same as in the space-like region. The  
solid curves which are calculated with Sudakov suppression describe the $q^2$  
dependence of  $q^4 G_m(q^2)$. The model $a)$ (bold solid curve) agrees with the  
experimental data rather well, while model $b)$ (thin solid curve) is below the  
data by a factor $\sim 1.5$. The dashed curves were calculated without Sudakov 
suppression. They correspond to rising $q^4 G_m (q^2)$ and are below the bold  
curves by a factor $\sim 1.2\div 1.5$ at $q^2=5$ $GeV^2$.

In section \ref{spin} two types of parametrization of
$T^{q\overline{q} \rightarrow N\overline{N}}_i$ were considered:
model i) and model ii). The $q^2$ dependence of the proton magnetic form
factor appeared to be not very sensitive to the choice of a model. However
the choice of the spin structure is important for separation of $G_e(q^2)$
and $G_m(q^2)$, and therefore for the description of $F_2(q^2)$. Our
predictions for $F_2$ are  sensitive to the choice of parameters $m$ and
$\left< {\bf k}_\perp^2\right>$ which enter eqs.(\ref{PauliFF})  and
(\ref{GmGeTilda}). For model i) when the non-spin-flip amplitude $D_1(s,{\bf
k}_\perp^2)\ {\bf 1}\cdot {\bf 1}$ dominates the value of $m$ is only
relevant.

In Fig. 5 we show the ratio $\ds \mu_p \frac{G_e}{G_m}$ in the space-like 
region. This ratio is slowly dependent on $Q^2$ in the region $\  0.5 \div 3.5$ 
$GeV^2$ and reaches the value of $\ds \mu_p \frac{G_e}{G_m}\approx 0.6\div
0.7$. Therefore for the ratio of electric to magnetic form factors  we have
$\ds \frac{G_e}{G_m}\approx 0.2 \div 0.25$. In our models i) and ii) the ratio
$\ds\frac{G_e}{G_m}$ is not  dependent on $q^2$:

\begin{eqnarray}\label{ratios}
\frac{G_e}{G_m} = \frac{m}{M} && \textrm{in model i) } , \nonumber \\ 
\frac{G_e}{G_m} = \frac{m}{M} +\frac{\left<{\bf k}_\perp^2\right>}{(m+M)^2}
\left( 2+  \frac{m}{M} \right) && \textrm{in model ii) } .
\end{eqnarray}

If the natural quark mass $m=0.22$ $GeV$ is chosen  model i) reproduces
the experimental data on $\ds \frac{G_e}{G_m}$ very well. Experimental data
on Pauli formfactor $F_2$ taken from  \cite{Bosted} suggest a slightly
larger value of $\ds \mu_p \frac{G_e}{G_m} \approx 0.7 \div 0.75$.  However, 
employing the same quark mass $m=0.22$ $GeV$ model i) describes quite well
the experimental data \cite{Bosted}.

Model ii) gives the ratio $G_e/G_m$  larger than model i) at the same values
of quark mass and $\left< {\bf k}_\perp^2 \right>$  (see  \ref{ratios}). As
it follows from (\ref{ratios}) to reproduce the experimental data for Pauli
form factor $F_2$  (or equivalently for $G_e/G_m$) in model ii) we should
take $m \simeq 0$, $\left< {\bf k}_\perp^2\right> \simeq 0.15$ $GeV^2$, the
choice seeming quite unnatural. Note, that if we put $\left< {\bf
k}_\perp^2\right>=0$ (in this  case the rapid quark moves in thedirection of
the hadron), model ii) provides for helicity conservation and the second
formula in (\ref{ratios}) reduces to the first one.

In fig 6. we present our results for Pauli form factor $F_2(Q^2)$ for two
parameterizations of Sudakov form factor. The solid line refers to the
parametrization of model $ a$), the thin line $-$ model $b$). The bold dashed
curve corresponds to the calculations with the scalar diquark exchange model
ii). For this curve the Sudakov form factor was taken in the $\alpha_S=const$
approximation (model $ a$). It is seen that even for small values of $\left<
{\bf k}_\perp^2\right>$ the magnitude of $F_2$ can not be reproduced by model
ii) but can be explained very well by model i).

Thus, the model developed in this paper reproduces experimental data for
nucleon form factors $G_m(q^2)$, $G_e(q^2)$ at negative (and positive in case
of $G_m$) $q^2$. Both parametrizations of Sudakov form factor result in the
suggested by experimental data $q^2$-dependence of the form factors. The
required normalization of Sach's form factor $G_m$ is achieved in model {\it
a)}. As for the spin structure of the quark-to-nucleon transition amplitudes,
model {\it a}), where $D_1(s,{\bf k}_\perp^2)$ dominates, correctly reproduces
spin effects even at modetate and small $Q^2$. In model ii) it is possible to
describe experimental data if unnaturally small values of $m$ and $\left< {\bf
k}_\perp^2\right>$ are employed.

To describe experimental data on $F_2$ quark mass was chosen at its natural
value of $ m = 0.22$ $GeV$. In model ii)  a combination of $m$ and $\left<{\bf
k}_\perp^2\right>$ accounts for the difference $\left[ G_e(q^2)-G_m(q^2)
\right]$ (see (\ref{GmGeTilda})). To provide an adequate description of
$F_2(q^2)$ in the space-like region one has to take $m\simeq 0\ GeV$ and
$\left< {\bf k}_\perp^2\right>=0.15$ $GeV^2$, a rather doubtful choice. Thus,
the comparison with experiment suggests, that the helicity conservation model
i) captures correctly spin effects even at moderate $q^2$.

The effective power dependence of $\left|G_m\left( s\right)\right|, \ \left|G_e \left(
s\right)\right| \sim s^{-2}$ suggested by experimental data can be explained
by two factors:\\  
\noindent
1) Power fall off of the $q \bar q \to N \bar N$ transition amplitude  $\ds
\frac{1}{s}\left| T^{q\overline{q}\rightarrow N\overline{N}}\left(s\right)
\right| \sim \left|  s \right|^{ -\left(1-2\alpha _B\left( 0\right) +
\alpha_M\left(0\right) \right)/2 }\sim \left|s\right|^{-5/4}$,

\noindent 2) Sudakov form factor,  which decreases with $s$ in the region
$10\ GeV^2 < s < 30 \ GeV^2$ as approximately $\left| S(s)\right| \sim\left|
s\right| ^{-3/4}$. Both parametrizations of Sudakov form factor $S^{(0)}$ and
$S^{(1)}$ have the same $s$-dependence in this region of $s$ $\vert
S(s)\vert\sim\left| s\right|^{-3/4}$.

The model reproduces the data at negative as well at positive $s$ very
well. However, while both $S^{(0)}(s)$ (\ref{Sudakov}) and $S^{(1)}(s)$
(\ref{OneLoopSudakov}) can describe $s$-dependence of $G_m(s)$, $F_2(s)$ and
$F_\pi(s)$ in the space-like region only $S^{(0)}(s)$ accounts for difference
in magnitudes of the form factors in the space- and time-like regions. The
ratios that $S^{(0)}(s)$, $S^{(1)}(s)$ predict (see (\ref{SudakovRatio}),
(\ref{SudakovRatioOneLoop})) are changed slightly because of the analytical
behavior of the square root that enters formulas (\ref{PionFFExpr}),
(\ref{GmGeTwoSpinors}), (\ref{PauliFF}) and (\ref{GmGeTilda}).

Pauli form factor $F_2\left( Q^2 \right)(Q^2)^{3}$ in the low $Q^2\approx 1\
GeV^2$-region  rises as $\sim Q^2$, while at asymptotically large $Q^2$
behaves as a constant. Threshold effects are taken into account by the factor
$\ds \left(\frac{q^2}{4M^2}-1\right)$.

In the case of pion form factor there is one important difference. In the
reaction $\gamma \rightarrow \pi^+ \pi^-$ the virtual photon produces a pair
of pseudoscalar particles in the final state, thus, the produced pions have
orbital momentum $l=1$. In this case amplitude the $A_z^{\gamma \to q\bar{q}}$
suppressed by the chirality conservation contributes, which results in  an
additional $\sim \vert q^2 \vert^{-1/2}$ factor in the pion form factor (see
(\ref{PionFFExpr})).

The asymptotic $q^2$-behavior of the transition $T$-amplitude governed by the
meson Regge trajectory $\alpha_M(t)$ has the  power fall off: $\ds
\frac{1}{q^2}\left| A^{q\bar q \to \pi^+ \pi^-}(q^2)\right| \sim \left| q^2 
\right|^{(\alpha_M(0)-1)/2}=  \left| q^2\right|^{-1/4}$. Additional  $\ds
\left( q^2 \right)^{-1/2}$ suppression arises due to the chirality
conservation. Therefore, in the absence of Sudakov form factor we have $\left|
F_\pi(q^2)\right| \sim |q^2|^{\alpha_M(0)/2-1}=|q^2|^{-3/4}$. Sudakov form
factor in the region of $(-q^2) \approx 5-10\ GeV^2$ behaves approximately as:
$S^{(0)}(q^2), S^{(1)}(q^2)  \sim (-q^2)^{-(0.3,\div 0.4)}$. Thus, the
function $Q^2 F_\pi(Q^2)$ decreases slowly  as $\left( Q^2 \right)^{-0.1-0.3}$
in this region of $q^2$.

We fixed parameters of Sudakov form factor in models $a)$ and $b)$ fitting
experimental data for the pion form factor: 

\noindent $\alpha_{S}=0.45$, $\omega^2=2.5 \ GeV^2$; $\Lambda_{QCD}=0.5 \
GeV$, $\mu_1 =1.5 \ GeV$, $\mu_2 = 1.5 \ GeV$. 

This choice makes it possible to get the desired behavior of the pion form
factor $q^2 F_{\pi}(q^2) \sim (q^2)^0$ in the region where experimental data
are available. Model $a)$ where the ratio $r_{ts}$ (\ref{SudakovRatio}) is
completely defined by the choice of $\alpha_{S}$ correctly reproduces the
magnitude of the pion form factor in the time-like region.

In fig. 7 we show our results for the pion form factor in the  space-like
region. Solid lines refer to calculations when Sudakov form factor is taken
into account: model $a)$ $-$ thick line, model $b)$ $-$ thin line. The remaining
dashed line shows the calculations without Sudakov form factor, which
results in a slow increase of $Q^2F_{\pi}(Q^2)$ in the considered domain of
$Q^2$.

Our results for the pion form factor in the time-like region are presented in
fig. 8. Here solid lines refer to parameterizations of Sudakov form factor in
the models $a)$ and $b)$. The thick and thin dashed lines represent
calculations without  Sudakov form factor. As it is the case in the space-like
region $q^2 F_{\pi}(q^2)$ is approximately constant in the domain of available
experimental data. In the model $a)$ the ratio of of the modulus of the pion
form factor in the time-like to space-like region is determined by the
analytical properties of Sudakov form factor and is equal to $\vert
F_{\pi}(q^2) \vert/ \vert F_{\pi}(-q^2) \vert \approx 2.5$ at $q^2 \approx 10
\ GeV^2$.

The difference in modulus of meson form factors in the space-like and
time-like regions in the framework of perturbative QCD with Sudakov effects
taken into account was discussed in ref. \cite{Gousset}. However, in the model
\cite{Gousset} the the difference of $\vert F_{\pi}(q^2) \vert$ and $\vert
F_{\pi} (-q^2) \vert$ is attributed mainly to the singularities
of the hard scattering amplitudes. In the model \cite{Geshkenbein} the pion
form factor is described using a phenomenological parameterization of its
imaginary part, and, thus, the different magnitude of $\vert F_{\pi} \vert$ is
provided for by $q^2$-dependence of the imaginary part of $F_{\pi}(q^2)$.
Note, that in our model such  corrections as discussed in \cite{Geshkenbein}
decrease as inverse powers of $q^2$  $\sim \ds \frac{M^2}{q^2}$, and if the
characteristic mass varies in the range of a few $GeV$ $M=1\div2\ GeV$ (as it
is the case in most hadronic reactions) this effect doesn't contribute
substantially already at $q^2 \sim 10 \div 20 \ GeV^2$. Moreover, if the
analytical behavior of Sudakov form factor is the major source of the
enhancement of hadronic form factors in the time-like region, the ratios
$r^{\pi}_{ts}=\vert F_{\pi}(q^2) \vert / \vert F_{\pi}(-q^2) \vert$ and
$r^{N}_{ts}=\vert F_{m}(q^2) \vert / \vert F_{m}(-q^2) \vert$  will decrease
very slowly as $q^2$ increases (see (\ref{SudakovRatio}) and
(\ref{SudakovRatioOneLoop})). For instance, in the model $a)$ $r^{(0)}_{ts}$
is constant and is approximately equal to $r^{(0)}_{ts} \approx 2.5$. In model
$b)$ $r^{(1)}_{ts} \approx 1.8$ at $q^2= 5 \ GeV^2$ and decreases to
$r^{(1)}_{ts}=1.35$ at $q^2 = 100 \ GeV^2$. Thus, measurements of
$r_{ts}(q^2)$ for pions and nucleons at large $q^2$ will resolve a fundamental
problem, whether or not the behavior of hadronic form factors at $q^2 \sim 10
\div 50 \ GeV^2$ is governed by perturbative or nonperturbative QCD. The
effect of difference of $q^2F_{\pi}(q^2)$ at $q^2>0$ and $q^2<0$ predicted in
our model also waits for experimental test.

\section{CONCLUSION.}\label{concl}

Employing QGSM we have investigated the form factors of pion and nucleon in
the space-like and time-like regions. Spin effects were taken into account by
introducing the respective quarks-to-hadrons $T^{q\bar q \to h \bar h}$ and
hadrons-to-quarks amplitudes $T^{h \bar h \to q\bar q}$, which allows one to
analyze spin effects of binary hadronic reactions and hadronic form factors.
In the framework of the developed model the experimental $q^2$-dependence of
the form factors $G_{m,e}\sim (q^2)^{-2}$, $F_{\pi} \sim (q^2)^{-1}$ up to
$q^2 \approx 100$ $GeV^2$ is explained in terms of the intercept of the
respective Regge trajectory and Sudakov form factor. The introduction of spin
variables provides the possibility to distinguish nucleon Sach's form factors
$G_m$, $G_e$ and calculate $F_{2}$. The pion form factor $F_{\pi}(q^2)$ has
additional suppression $\sim 1/\sqrt{q^2}$ in the model due to the approximate
chirality conservation in the process $\gamma \to q \bar q$. QGSM predicts
asymptotic ratios of form factors $F_{\pi\pi}/F_{\pi \omega}$, $F_{\pi
\pi}/F_{\pi \rho} \sim 1/\sqrt{q^2}$, while in perturbative QCD those ratios
rise linearly as $q^2$ increases. The obtained expressions for hadronic form
factors are analytical in the $q^2$ complex plained and, thus, can be
continued from the time-like to space-like region. The difference of the
modulus of $G_{m,e}(s)$ and $F_{\pi}(s)$ at positive and negative $s$ is
mainly related to analytical properties of the doubly logarithmic term in the
exponent of Sudakov form factor. Sudakov form factor in KGSM leads to hadronic
form factors which decrease faster than any power of $\ds {1}\sqrt{q^2}$ at
very large $\vert q^2\vert$. However, this exponential suppression doesn't
contribute strongly in the domain of currently accessible $q^2$. As a result
nonperturbative effects predicted in the framework of KGSM may be dominant up
to $q^2 \approx 10^2 \ GeV^2$. 

\vspace{1.5cm}  
The authors are grateful to R. Baldini, K.G. Boreskov, B.V. Geshkenbein,
O.V. Kancheli and Yu.A. Simonov for useful discussion and to C.F. Perdrisat
for sending us experimental data on $G_e / G_m$ before publication. This work
was partially supported by the grants RFFI 98-02-17463 and NATO
OUTR.LG.971390.

\newpage

{\Large {\bf Figure captions}}

\vspace{1cm} 

\noindent Fig. 1. Planar diagrams for binary reactions
 {\bf a)} $ \pi^0
\pi^0 \to \pi^- \pi^+$, {\bf b)} $\pi \bar \pi \to N \bar
 N$, {\bf c)} $N
\bar N \to N \bar N$.

\vspace{0.5cm}

\noindent Fig. 2. Planar diagrams for reactions {\bf a)} $\gamma \to \pi^+
\pi^-$, {\bf b)} $\gamma \to N \bar N$.

\vspace{0.5cm}

\noindent Fig. 3. Proton magnetic form factor $G_m(Q^2)$ in the space-like
region as function of $Q^2=-q^2$. Thick and thin solid lines refer to  QGSM
calculations  with parameterizations (\ref{Sudakov}) and
(\ref{OneLoopSudakov}) of Sudakov form factor. Thick and thin dashed line are
obtaind from the respective solid lines by dividing by Sudakov form factor.
The experimental data are taken from \cite{Arnold}.

\vspace{0.5cm}

\noindent Fig. 4. Proton magnetic form factor $G_m(q^2)$ in the time-like
region as a function of $q^2$. Thick and thin solid lines refer to to QGSM
calculations  with parameterizations (\ref{Sudakov}) and
(\ref{OneLoopSudakov}) of Sudakov form factor. Thick and thin dashed line are
obtained from the respective solid lines by dividing them by Sudakov form
factor. The experimental data are taken from \cite{ Fermilab,Bisello1}.

\vspace{0.5cm}

\noindent Fig. 5. The ratio of electric to magnetic proton form factors
$\mu_p G_e(Q^2)/ G_m(Q^2)$ in the space-like region as a function $Q^2
=-q^2$. The calculations were done using model i) for the spin structure of
the $q \bar q \to N \bar N $ amplitude with quark mass taken $m=0.22\ GeV$.
The experimental data are taken from \cite{Perdrisat}.

\vspace{0.5cm}

\noindent Fig. 6. Pauli proton form factor $F_2(Q^2)$ in the space-like
region as a function of $Q^2 = - q^2$. Solid lines are the results of our
calculations within model i) for the $A^{q\bar q \to N \bar N}$ amplitude;
thick line represents calculations with parameterization (\ref{Sudakov}) of
Sudakov form factor, thin line $-$ with parametrization
(\ref{OneLoopSudakov}). The respective dashed lines are the calculations
without Sudakov form factor. The lower bold dashed line corresponds to
parameterization ii) of spin amplitudes and (\ref{Sudakov}) of Sudakov form
factor. Parameters $m$ and $\left< {\bf k}^2_\perp \right>$ were taken equal
to $m=0.22 \ GeV$, $\left< {\bf k}^2_\perp \right> = 0.2 \ GeV^2$.
Experimental data are taken from \cite{Bosted}.   \vspace{0.5cm}

\noindent Fig. 7. Pion form factor $F_{\pi}(Q^2)$ in the space-like region as
function of $Q^2 =-q^2$. Bold and thin solid lines refer to calculations with
parametrizations (\ref{Sudakov}) and (\ref{OneLoopSudakov}) for Sudakov form
factor respectively. Bold and thin dashed lines refer to QGSM calculations
without Sudakov form factor. These curves are obtained from the respective
solid ones  dividing them by Sudakov form factor. Experimental data are taken
from \cite{Bebek}.

\vspace{0.5cm}

\noindent Fig. 8. Pion form factor $F_{\pi}(q^2)$ in the time-like region as
function of $Q^2 =-q^2$. Thick and thin solid lines refer to calculations 
with parameterizations ((\ref{Sudakov}) and (\ref{OneLoopSudakov}) for
Sudakov form factor respectively. Bold and thin dashed lines refer to QGSM
calculations without Sudakov form factor. These curves are obtained from the
respective solid ones dividing them by Sudakov form factor. Experimental data
are taken from \cite{ BiselloPion,Jpsi}.

\newpage

\begin{figure}[htbp]\label{RectangleDiagramms}
\centerline{\epsfbox{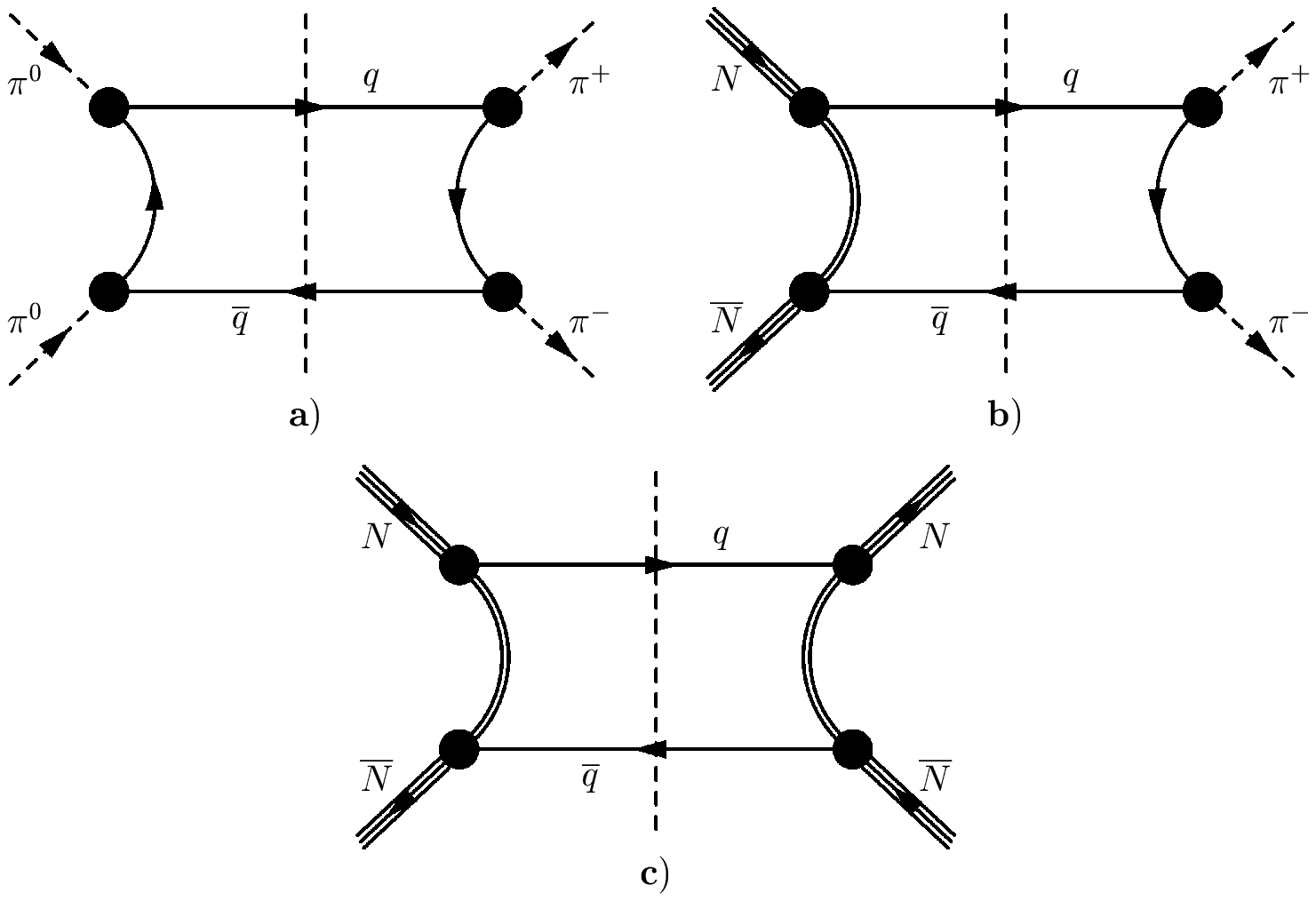}}
\end{figure}
\newpage


\begin{figure}[htbp]\label{TriangleDiagramms}
\centerline{\epsfbox{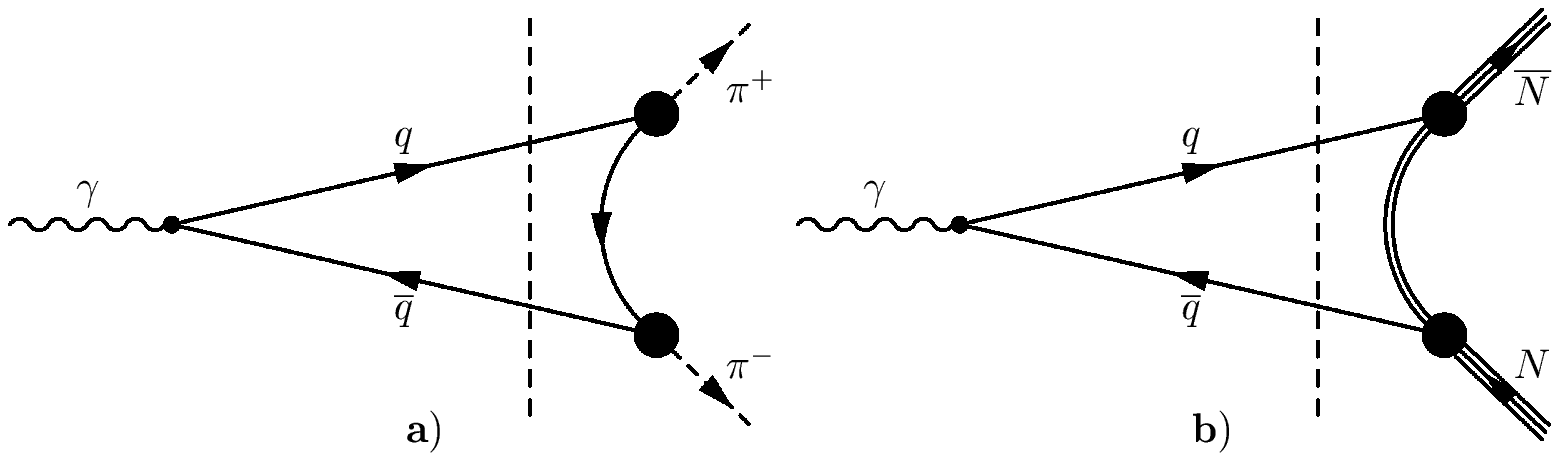}}
\end{figure}

\newpage

\begin{figure}[htbp]\label{Gmsl}
\centerline{\psfig{figure=gmsl.epsi,width=18cm}}
\end{figure}

\newpage

\begin{figure}[htbp]\label{Gmtl}
\centerline{\psfig{figure=gmtl.epsi,width=18cm}}
\end{figure}

\newpage

\begin{figure}[htbp]\label{gegmsl} 
\centerline{\psfig{figure=gegmsl.epsi,width=18cm}}
\end{figure}

\newpage

\begin{figure}[htbp]\label{F2sl} 
\centerline{\psfig{figure=f2sl.epsi,width=18cm}}
\end{figure}

\newpage

\begin{figure}[htbp]\label{Fpsl}
\centerline{\psfig{figure=fpsl2.epsi,width=18cm}}
\end{figure}

\newpage

\begin{figure}[htbp]\label{Fptl}
\centerline{\psfig{figure=fptl2.epsi,width=18cm}}
\end{figure}

\end{document}